# Loops, Loop Sequences and Loop Surfaces in Statistical Geodesics, and Quantum Information

Aalok*

Department of Physics, University of Rajasthan, Jaipur 302004, India;
and Jaipur Engineering College and Research Centre (JECRC), Jaipur 303905, India.

**Abstract**

The locus of probability flow in Quantum Mechanics and information is explored. We explore *loops*, *loop sequences* and *loop surfaces* in the statistical geodesics. Having known about the loop character of the statistical geodesics in probability space, we discuss wider physical aspects of consequences of the geometry of the probabilities in the form of *loop sequences* and *loop surfaces*. These are interpreted as the *locus of probability flow*. A deeper philosophical observation reveals that these loops, loop sequences and loop surfaces are the information grids. These investigations in many ways could be significant not only in the research in Statistics, but also in Mathematics, and Physics in general, and entropy, gravity and information theory in specific.

**Keywords**: **Quantum state space; Projective Hilbert space;** Kähler Manifold; **Statistical distance; Statistical geodesic; Probability and Geometry; Loop Quantum Gravity.**

.

___________________________________

*\*E-mail*: aalok@uniraj.ernet.in

# 1. Introduction

Having inspired by recent studies [1] of statistical geodesics and distances in the form of probability flow and loops, we further explore the geometry of probability space. The importance of statistical geometry, which is essentially geometry of probability, was put on a profound footing by Brody and Hughston [2, 3] in the context of geometry of quantum mechanics. This is also found to be important in the case of statistical inference and decision making as deliberated in Arima and Nagaoka [4, 5] and Amari [5].

The statistical distance function and the notion of metricity, was given a concrete shape using differential geometry by Rao [6] and Rao in Amari [5].

A statistical model $\mathcal{M}$ is a family of probability distributions, characterized by a set of continuous parameters known as parameter space. One can regard $\mathcal{M}$ as a submanifold of the unit shere $\mathcal{S}$ in a real Hilbert space $\mathcal{H}$. Therefore, $\mathcal{H}$ embodies the 'state space' of the probability distributions, and the geometry of the given statistical model can be described in terms of the embedding of $\mathcal{M}$ in $\mathcal{S}$. The geometry in question is characterized by a natural Riemannian metric known as Fisher-Rao metric as appeared in Brody and Hughston [2, 3] and in Arima and Nagaoka [4] and Amari [5]. There are a number of distinct geometrical formulations of classical statistical theory, corresponding, for example, to the various–embeddings of Amari and Nagaoka [4], and Amari [5], but one among these is singled out on account of its close relation to quantum theory: the geometry of square-root density functions. This geometry is special because of the way it singles out the Levi-Civita connection on statistical sub-manifolds. And as a result, we are led to consider classical statistics in the language of real Hilbert-space geometry as given by Brody and Hughston [3].

Thus the studies and investigations of geodesics and distances in Statistics have come a long way as evident from Brody and Hughston [2, 3], Arima Nagaoka [4] and Amari [5],

Bengtsson [7, 8], Braunstein and Caves [9], and Braunstein [10]. Fisher information is related to the 'velocity' along the given curve in Hilbert space. This is a result that has profound links with analogous constructions in quantum mechanics is prominently described by Anandan and Aharonov [11].

It is worth mentioning here the importance of the geometrization of Quantum Mechanics discussed by Kibble [12], which pointed out that the Schrödinger evolution can be regarded as Hamiltonian flow on $\mathcal{H}$. We wish to recast the same spirit in terms of probability flow in $\mathcal{P}$, and in Probability space as described by Aalok [1].

We continue to explore what is emphasized by Aalok [1] for the Faraday lines and loops as the *locus of probability flow*. It is interesting to notice that the seminal work of Fraday [13] is manifested in the contemporary Physics again and again in one context or the other. Having known about the loop character of the statistical geodesics in probability space [1], we explore the geometry of the probability space in the form of *loop sequences* and *loop surfaces*. This in many ways could be significant not only in the ongoing research in Statistics and Mathematics, but also in Physics in general and *Loop Quantum Gravity* (LQG) in specific. The idea of loops and loop surfaces finds a prominent place in theory of Loop Quantum Gravity [14-16]. The hypothesis of graphs and loops even in the absence of physical objects or dynamics makes sense in terms of probability and probable paths. This is also motivation for us to explore these ideas in probability space [1].

This paper will generate a worthwhile debate and will follow a trail of explorations on statistical geodesics and distances, we believe. Also, we discuss some of the important information measures such as Shannon entropy, Rényi entropy, Tsallis entropy, Abe entropy, Landsberg- Vedral entropy, Kaniadakis entropy and Sharma- Mittal entropy, that are useful for the description of complex systems and analyze their geometric character in the light of present discussion.

## 2. The Geometry of the Probability Space

Having inspired by the idea of probability flow and loops, we explore the geometry of probability space. In Statistics, the probability distribution for the outcome of a physical process is given by $n+1$ real numbers $p_i$ such that

$$p_i \geq 0, \text{ and } \sum_{i=0}^{n} p_i = 1 . \tag{1}$$

The geometry of the $n$- sphere is almost manifest in [4], and we find it such that

$$\xi_i \equiv \sqrt{p_i} \implies \sum_{i=0}^{n} \xi_i^2 = 1 . \tag{2}$$

We see here that the space of all probability distributions lies on a sphere embedded in a flat space as given in [4], and therefore it carries the natural metric of the sphere, namely:

$$ds^2 = \sum_{i=0}^{n} d\xi_i d\xi_i = \frac{1}{4} \sum_{i=0}^{n} \frac{dp_i dp_i}{p_i} . \tag{3}$$

In Statistics this is known as the Bhattacharya metric or Fisher-Rao metric [2-5, 7-10]. It enables us to define the geodesic distance between two arbitrary probability distributions; if we consider the case where there are only two possible outcomes then the geodesic distance between two probability distributions $(1-p_1, p_1)$ and $(1-p_2, p_2)$ as:

$$\cos d = \sqrt{p_1 p_2} + \sqrt{(1-p_1)(1-p_2)} . \tag{4}$$

Therefore distance $d$ is given by:

$$d = \cos^{-1}(\sqrt{p_1 p_2} + \sqrt{(1-p_1)(1-p_2)}) . \tag{5}$$

Using the identity of inverse circular functions we find that- the expression of geodesic in probability space is equivalent to

$$d = \cos^{-1}(\sqrt{p_1}) - \cos^{-1}(\sqrt{p_2}) . \tag{6}$$

The geometry of these geodesic expressions has already been explored [1].

The graphical plot of the geodesic

$$d = \cos^{-1}(\sqrt{p_1}) - \cos^{-1}(\sqrt{p_2}) ; \qquad (7)$$

appears as follows:

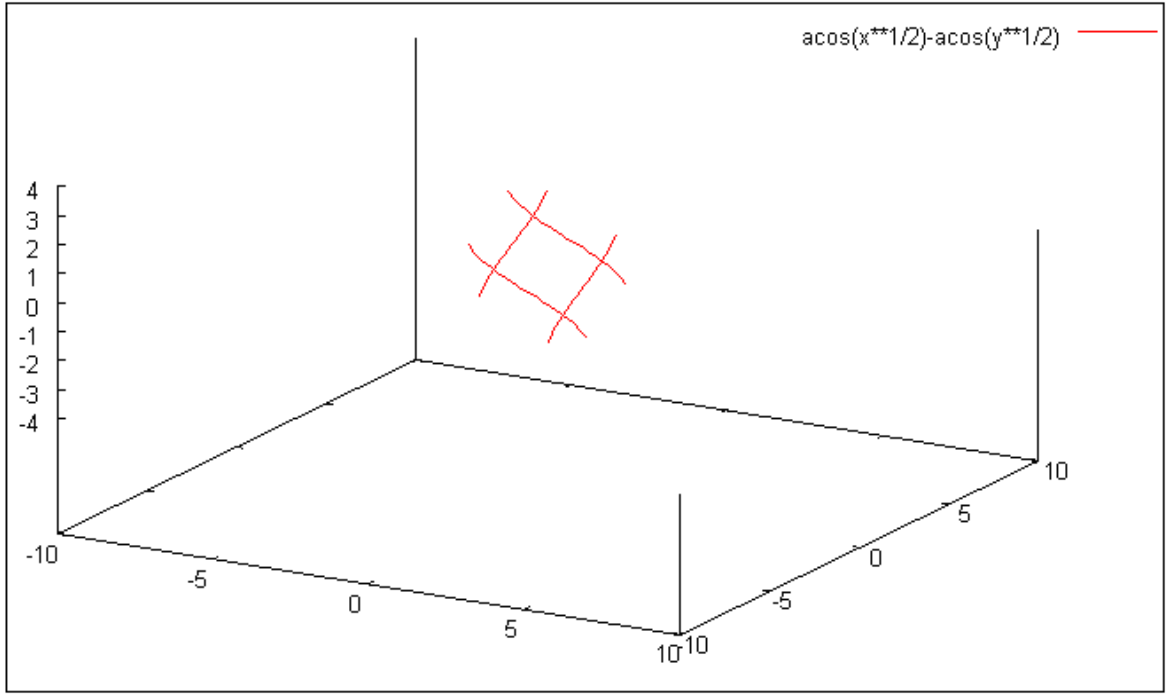

**Fig. 1**: Graph representing geodesic $d = \cos^{-1}(\sqrt{p_1}) - \cos^{-1}(\sqrt{p_2})$ in probability space.

This is precisely a loop form. The joints of the loop imply the possibilities of further loop attachments with this loop and so on. And ultimately that may result into a mesh or grid. Also, it is interesting to note that abstract graph structure is not directed one. It is only in a specific physical framework that these graphs appear to be directed.

Also, the statistical distance for binomial distribution of probabilities is given [6] by:

$$d = \sin^{-1}(\sqrt{p_1}) - \sin^{-1}(\sqrt{p_2}) . \qquad (8)$$

If we plot the expression in eq. (8), this too is a loop form in a different plane as follows.

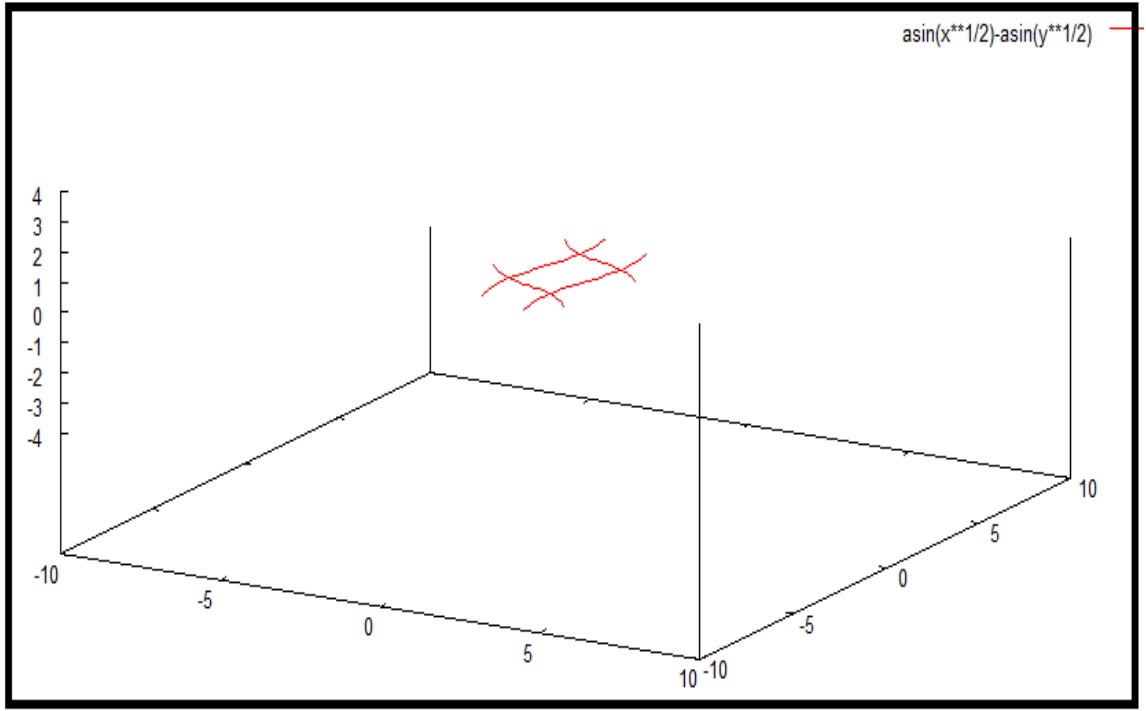

**Fig. 2**: Graph representing geodesic $d = \sin^{-1}(\sqrt{p_1}) - \sin^{-1}(\sqrt{p_2})$ in probability space.

We have an interpretation of this expression that even if we do not consider any separation between two probabilities $p_1$ and $p_2$, there exists a geodesic corresponding to each of the probabilities $p_1$ and $p_2$ separately. That is if one can afford to drop any of the two terms, one finds that

$$d = \cos^{-1}(\sqrt{p_1}) ; \tag{9}$$

or a geodesic with the second part as

$$d = \cos^{-1}(\sqrt{p_2}) , \tag{10}$$

The corresponding geometry could be found as the following *loop sequence*:

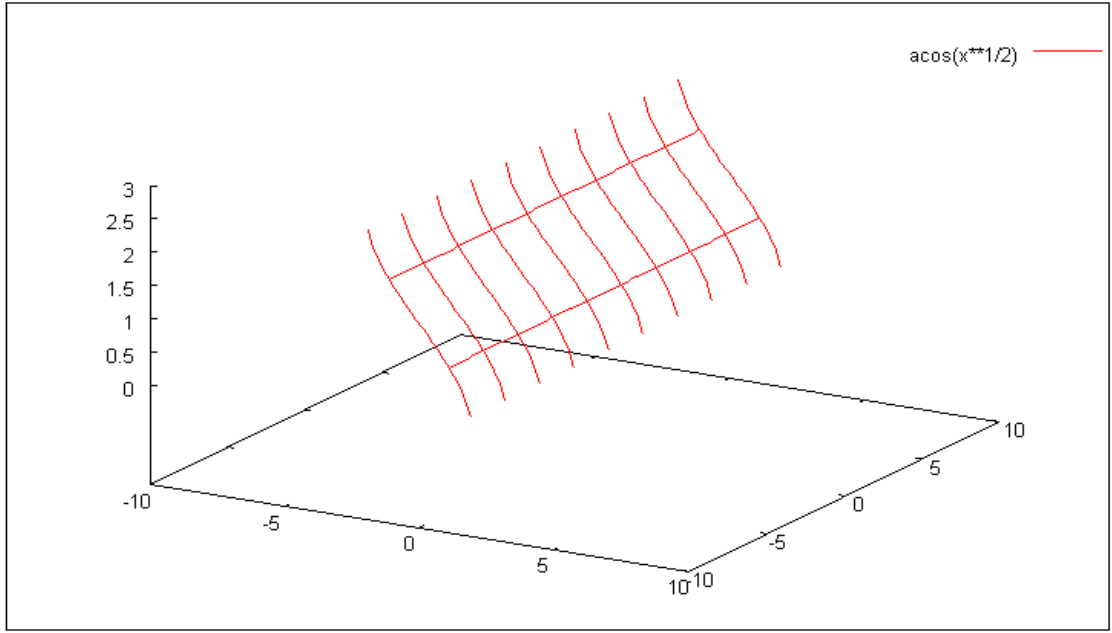

**Fig. 3:** Graph representing $d = \cos^{-1}(\sqrt{p_1})$ in probability space: A *loop sequence*.

A philosophically relevant observation crops up that even in the absence of probability $p_2$, geodesic with probability $p_1$ alone exists and manifests in the form of geometry of *loop sequence*. If we overlay the geometry of the two expressions in the equations (9) and (10), we find the resulting graph as two independent *loop sequences* as follows:

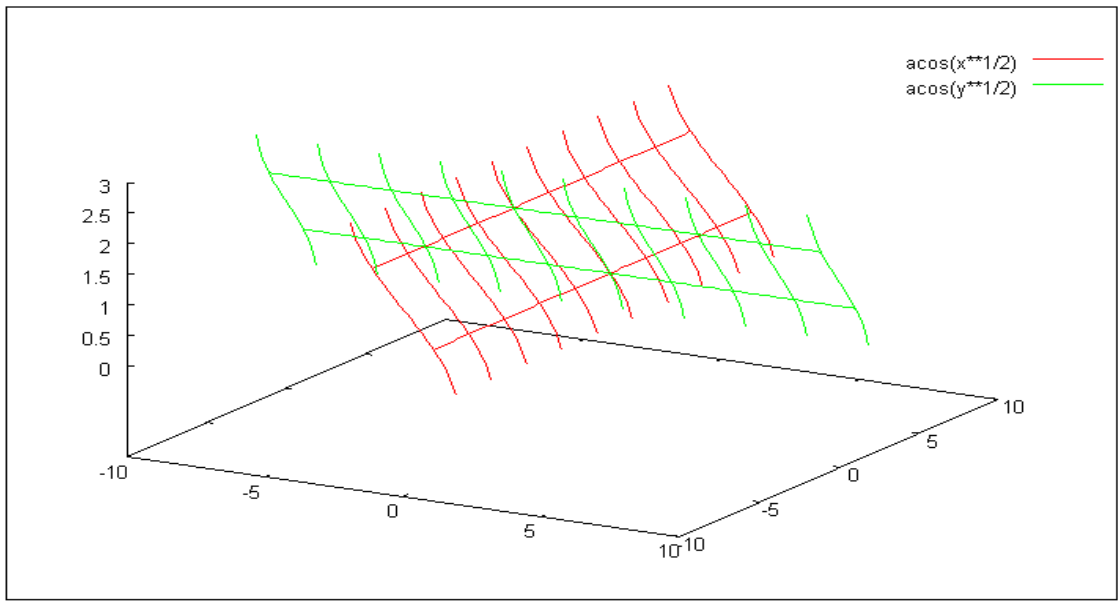

**Fig. 4:** Overlaying two independent *loop sequences* $\cos^{-1}(\sqrt{p_1})$; and $\cos^{-1}(\sqrt{p_2})$:
Co-existence of two independent geodesics in the probability space.

The overlaying of the two loop sequences simply implies the independent existence of the two geodesics in the probability space. The expression of statistical geodesic is often given in an alternative form as:

$$d = \cos^{-1}(\sqrt{p_1}\sqrt{p_2}). \quad (11)$$

If we plot geometry of this expression, we finally obtain a nice sprawling grid or *loop surface* as given in the Figure (5).

In case one finds it difficult to plot the above equation, one can extrapolate it as follows.

Take log of the expression $\cos d = (\sqrt{p_1}\sqrt{p_2})$, which appears as:

$$\log(\cos d) = \log(\sqrt{p_1}\sqrt{p_2}).$$

Or $\log(\cos d) = \log(\sqrt{p_1}) + \log(\sqrt{p_2})$,

or $\cos d = \exp[1/2\{\log(p_1) + \log(p_2)\}]$,

or $d = \cos^{-1}(\exp[1/2\{\log(p_1) + \log(p_2)\}])$. (12)

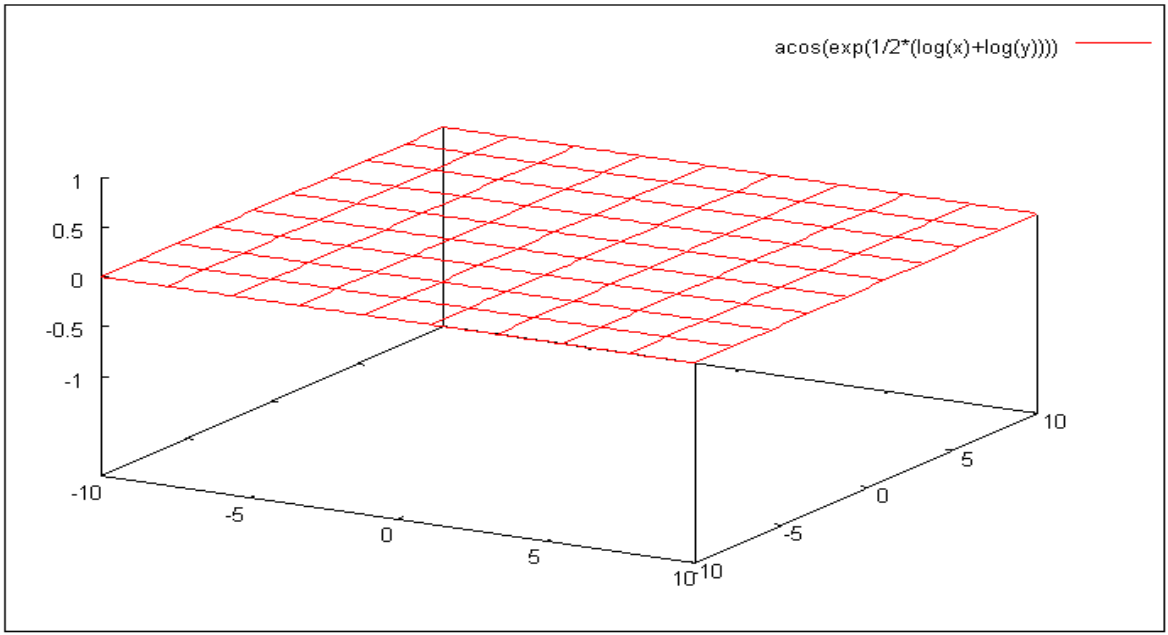

**Fig. 5:** Graph representing $d = \cos^{-1}(\sqrt{p_1 p_2})$ in probability space: A *loop surface*.

Again this makes us notice philosophically relevant observations. The expression of geodesic in equation (5), if we drop the second term then results into equation (11). The second part in the equation (5) has terms $\sqrt{(1-p_1)}$ and $\sqrt{(1-p_2)}$ that stand for probabilities of not happening of $p_1$ and $p_2$ respectively. Thus it is pertinent to notice that if we include the complementary probabilities $1-p_1$ and $1-p_2$ along with probabilities $p_1$ and $p_2$ in the information measure, we get more definitive picture in the form of a single loop only. And if we do not include the complementary probabilities, we can only have a loop sequence or a loop surface with vast extension of probabilities. Now an important question might arise as: What does this mesh or grid represent? Of course, it is *probability distribution*. And, in terms of information, one may call it- *spread of the information* or at best the *information network*. A deeper philosophical observation reveals that these loops, loop sequences and loop surfaces are in fact information grids.

## 3. The Cases of Physical Importance

It is not that the present discussion pertains only to abstract aspects of mathematical and statistical importance. We essentially draw inspiration from the cases of interest for physicists.

**(i) Geometry of Shannon entropy in quantum mechanical formulation**

In the absence of any explicit information the uniform probability distribution is the best bet, and then the problem becomes how to upgrade this suggestion when the results of the experiments are coming in. This problem occurs in statistical mechanics, and the method used there is to maximize the Shannon entropy:

$$S = -\sum_i p_i \ln p_i ; \qquad (13)$$

Subject to the constraints

$$\sum_{i=1}^{n} p_i = 1, \text{ and } \langle a \rangle \equiv \sum_{i=1}^{n} a_i p_i = Const. \tag{14}$$

The second constraint is to be thought of as the result of some experiment. The Shannon entropy has some unique features, which need not be discussed here. Statistically speaking this is an optimization problem. We explore the geometry of the entropy function widely used for example in Bengtsson [7, 8], and in Rovelli [14-16], is given by

$$S = -\sum_i p_i \ln p_i . \tag{15}$$

In fact $S$ is a geometric expression of entropy. Up to an arbitrary multiplicative constant, one can easily show [17, 18] that this is the entropic form that easily satisfies all four Khinchin axioms that it follows uniquely (up to a multiplicative constant) from these postulates. If we plot Shannon's entropy, we get a nice loop surface as given in the Fig. (6).

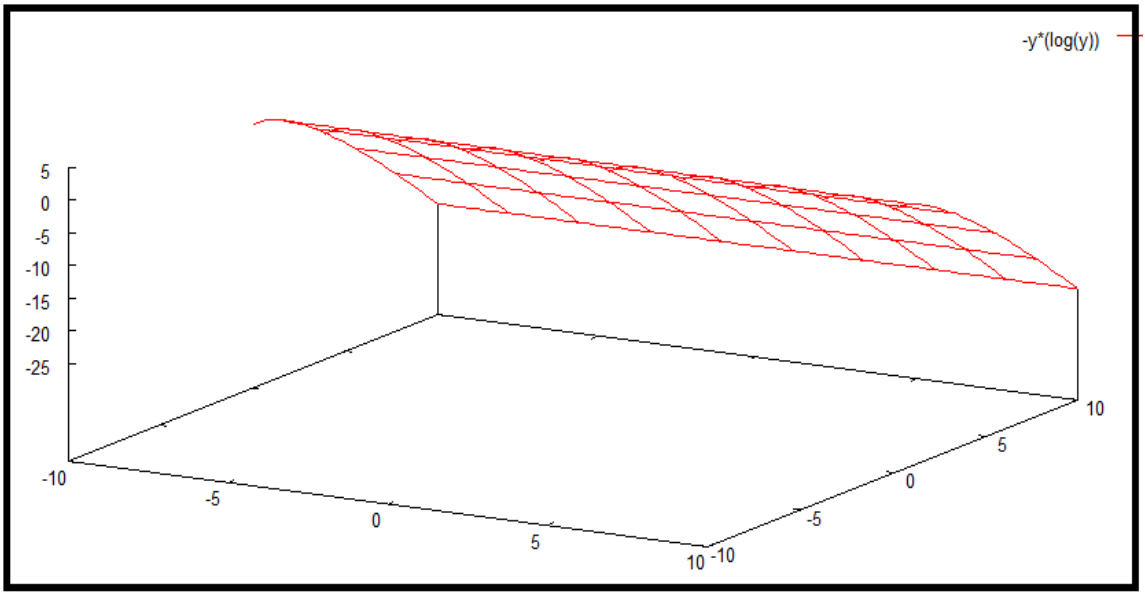

**Fig. 6:** Graph representing Shannon entropy $S = -\sum_i p_i \ln p_i$ in the probability space.

It is interesting to note that this surface is curved too. The reason behind this is very clear. The curvature of this statistical metric stems from the fact that the statistical fluctuations are much smaller when we are close to the pure state that is close to the edges of the simplex.

Now yet another question crops up as: Is this surface same as the much talked about trapped surface in the black hole description? The curvature or the cavity of this surface is inward and may be this is the reason why any information does not come out of Black-Holes. It needs further investigation. For the uniform distribution $p_i = \frac{1}{W}$, the Shannon entropy takes on its maximum value:

$$S = k \ln W .\tag{16}$$

This is famous Boltzmann's entropy formula.

Maximization of the Shannon entropy subject to suitable constraints leads to ordinary statistical mechanics [17-19]. In thermodynamic equilibrium, the Shannon entropy can be identified as the 'physical' entropy of the system, with the usual thermodynamic relations. Generally, the Shannon entropy has an enormous range of applications not only in equilibrium Statistical Mechanics but also in Coding Theory, Computer Science etc. it is easy to verify that Shannon entropy is a concave function of the probabilities $p_i$, which is an important formulate of Statistical Mechanics.

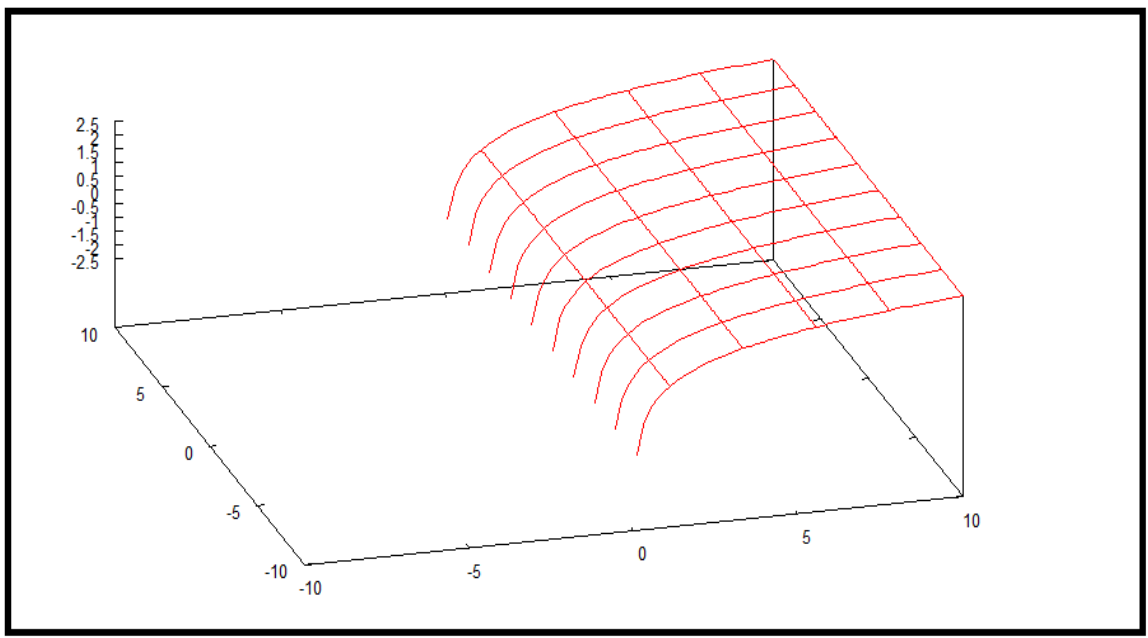

**Fig. 7:** Graph representing Boltzmann entropy $S = k \ln W$.

We now discuss some more general information measures.

**(ii) Geometry of the Rényi entropies**

The Rényi information measure is important for the characterization of multi-fractal sets that is fractals with a probability measure on their support as well as for certain applications in computer Science [17-19].

With the consideration that the entropy of independent systems should be additive, we encounter an information measure that is called- Rényi entropy [20]. For an arbitrary real parameter $q$, Rényi entropy is defined as:

$$S_q^R = \frac{1}{q-1} \ln \sum_i p_i^q . \qquad (17)$$

The geometrical plot of this entropic function appears as follows:

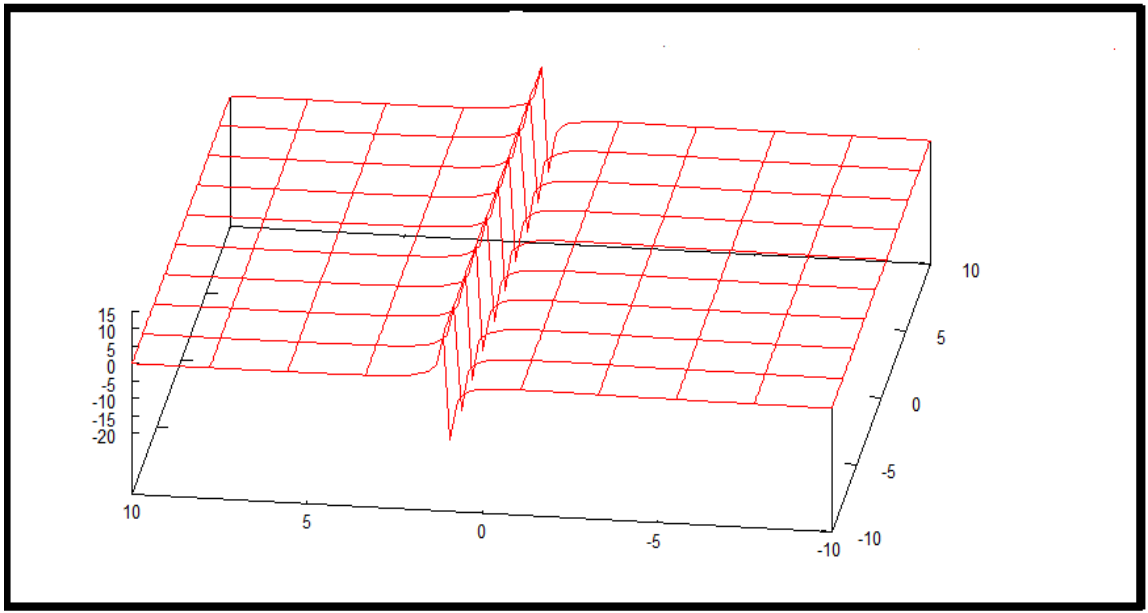

**Fig. 8:** Graph representing Rényi entropy

The summation is over all events $i$ with $p_i \neq 0$. For $q \to 1$ Rényi entropy reduces to the Shannon entropy:

$$\underset{q\to 1}{Lim}\, S_q^R = S, \quad (18)$$

It can be easily derived by setting $q = 1+\varepsilon$ and doing a perturbative expansion in the small parameter $\varepsilon$ in eq. (17).

### (iii) Geometry of the Tsallis entropies

The Tsallis entropy is given by the following expression [21]:

$$S_q^T = \frac{1}{q-1}\left(1 - \sum_i^W p_i^q\right). \quad (19)$$

The geometrical plot of this entropic function appears as follows:

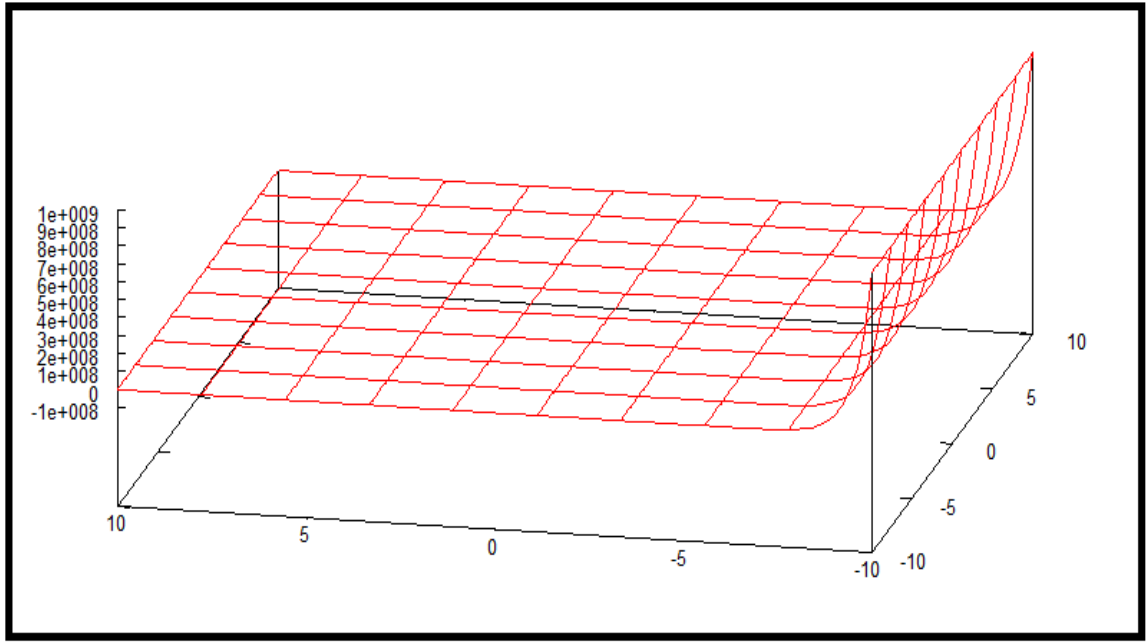

**Fig. 9:** Graph representing Tsallis entropy

It was Tsalli who in his seminal papers [21, 22] had suggested a generalization of Statistical Mechanics using these entropic forms. Tsallis entropies are different from Rényi entropies, as it does not have logarithmic term any more. A relation between Rényi and the Tsallis entropy is easily derived by writing

$$\sum_i p_i^q = 1 - (q-1)S_q^T = e^{(q-1)S_q^R}. \quad (20)$$

However, the Tsallis entropy is a (monotonous) function of the Rényi entropy, so any maxima of Tsallis entropy imply maxima of Rényi entropy and *vice-versa*. But still, Tsallis entropies have many distinguished properties that make them a better (suitable) candidate for generalization of Statistical Mechanics than Rényi entropies. One such property is concavity. One can easily verify that-

$$\frac{\partial}{\partial p_i} S_q^T = \frac{-q}{q-1} p_i^{q-1} ; \text{ and} \tag{21}$$

$$\frac{\partial^2}{\partial p_i \partial p_j} S_q^T = -q p_i^{q-2} \delta_{ij}. \tag{22}$$

This means that, as a sum of concave functions $S_q^T$ is concave for all $q > 0$ (concave for all $q < 0$). This property does not hold for Rényi entropies.

The Tsallis entropy also reduces to Shannon entropy. As a special case of $q \to 1$, Tsallis entropy imply:

$$S_1^T = \lim_{q \to 1} S_q^T = S \ (shannon). \tag{23}$$

As a good information measure the Tsallis entropy assume their extremum for the uniform distribution $p_i = \frac{1}{W} \forall i$. This extremum is given by

$$S_q^T = \frac{W^{1-q} - 1}{1 - q}; \tag{24}$$

Which, in the limit $q \to 1$, reduces to Boltzmann's celebrated formula: $S = k \ln W$.

**(iv) Geometry of Landsberg- Vedral entropy**

The measure with the following expression:

$$S_q^L = \frac{1}{q-1} \left( \frac{1}{\sum_{i=1}^{W} p_i^q} - 1 \right), \tag{25}$$

was explored by Landsberg and Vedral [23]. We immediately realize that the Landsberg-Vedral entropy is related to the Tsallis entropy $S_q^T$ by

$$S_q^L = \frac{S_q^T}{\sum_{i=1}^{W} p_i^q}, \tag{26}$$

And hence $S_q^L$ is sometimes also called normalized Tsallis entropy. $S_q^T$ also contains the Shannon entropy as a special case

$$\underset{q \to 1}{Lt}\, S_q^L = S_1. \tag{27}$$

The geometrical plot of this entropic function appears as follows:

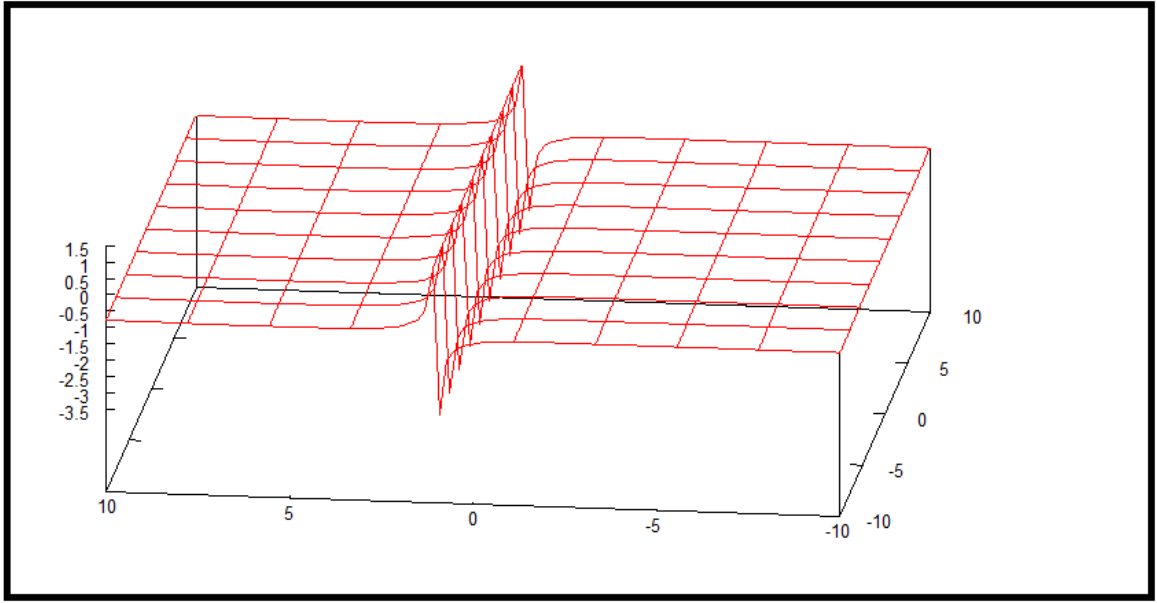

**Fig. 10:** Graph representing Landsberg- Vedral entropy

**(v) Geometry of Abe entropy**

Abe [24] introduced a kind of symmetric modification of Tsallis entropy, which is invariant under the exchange $q \leftrightarrow q^{-1}$. This is given by:

$$S_q^{Abe} = -\sum_i \left( \frac{p_i^q - p_i^{q^{-1}}}{q - q^{-1}} \right). \tag{28}$$

This symmetric choice in $q$ and $q^{-1}$ is inspired by the theory of quantum groups which often exhibits invariance under the 'duality transformation' $q \leftrightarrow q^{-1}$. The geometrical plot of this entropic function appears as follows:

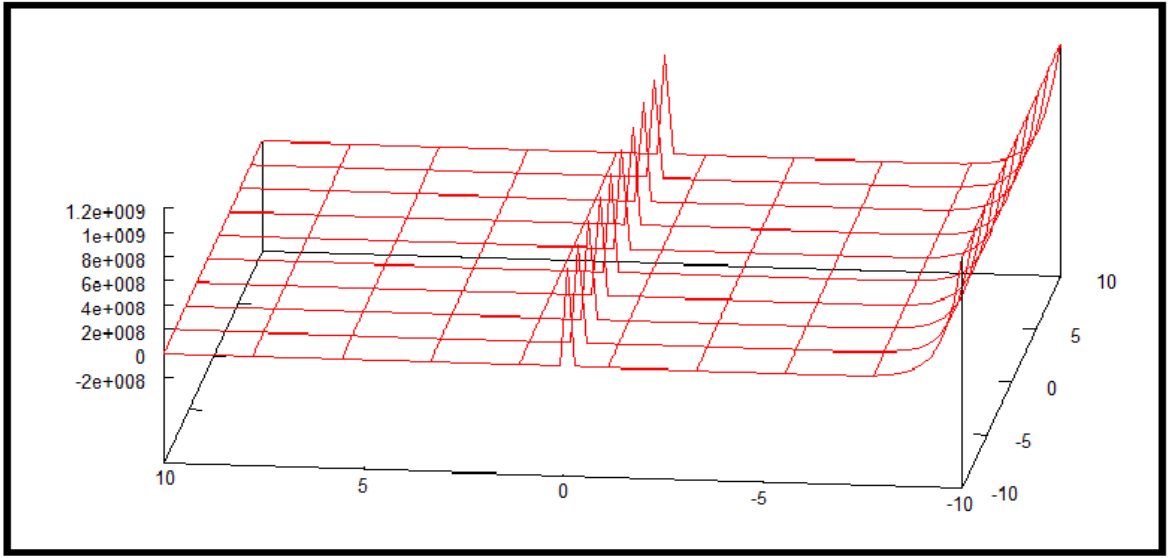

**Fig. 11:** Graph representing Abe entropy

Apparently, Abe entropy has singularity at $q = 0$. If we wish to avoid this singularity and reexamine it, we substitute $q \to q + \varepsilon$ in the limits when $\varepsilon$ is infinitesimally small but not zero, the Abe entropy appears as follows in **Fig**. 12.

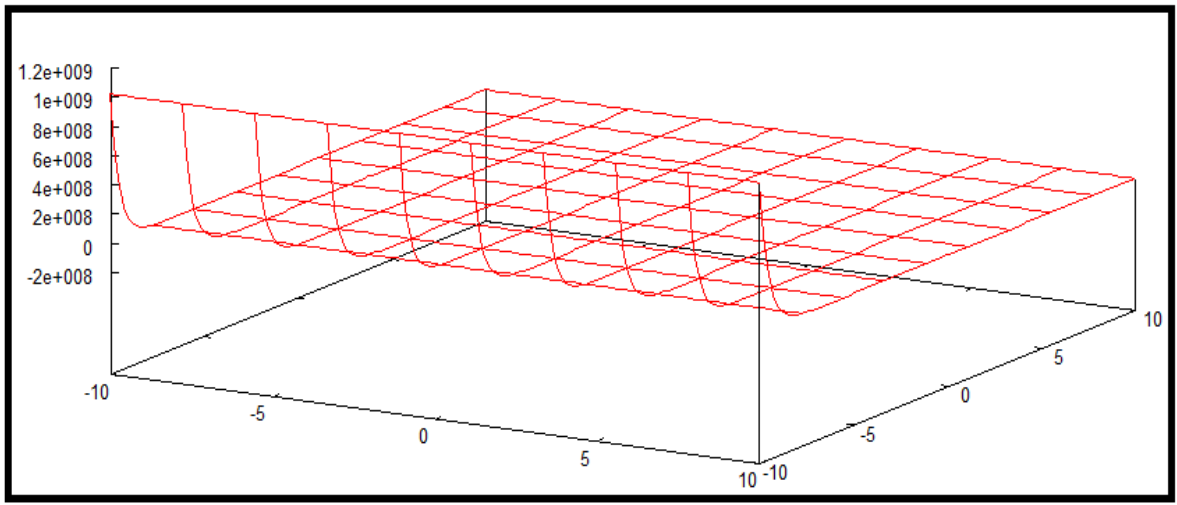

**Fig. 12:** Graph representing Abe entropy free from singularity at $q = 0$.

This geometry is similar to the geometry of Tsallis entropy.

**(vi) Geometry of Kaniadakis entropy**

The Kaniadakis entropy or $\kappa$ entropy is defined by the following expression [25]:

$$S_\kappa = \sum_i \left( \frac{p_i^{1+\kappa} - p_i^{1-\kappa}}{2\kappa} \right). \tag{29}$$

It is pertinent to mention that it is a kind of Shannon entropy, which reduces to the original Shannon entropy for $\kappa = 0$. We also note that for small $\kappa$, and by writing $q = 1 + \kappa$ and $q^{-1} \approx 1 - \kappa$, the Kaniadakis entropy approaches the Abe entropy. Kaniadakis was motivated to introduce this entropic form by special relativity wherein the relativistic sum of two velocities of particles of mass $m$ satisfies as similar relation as the Kaniadakis entropy does, identifying $\kappa = \frac{1}{mc}$. Kaniadakis entropies are also concave and Lesche stable. The geometrical plot of this entropic function appears as follows:

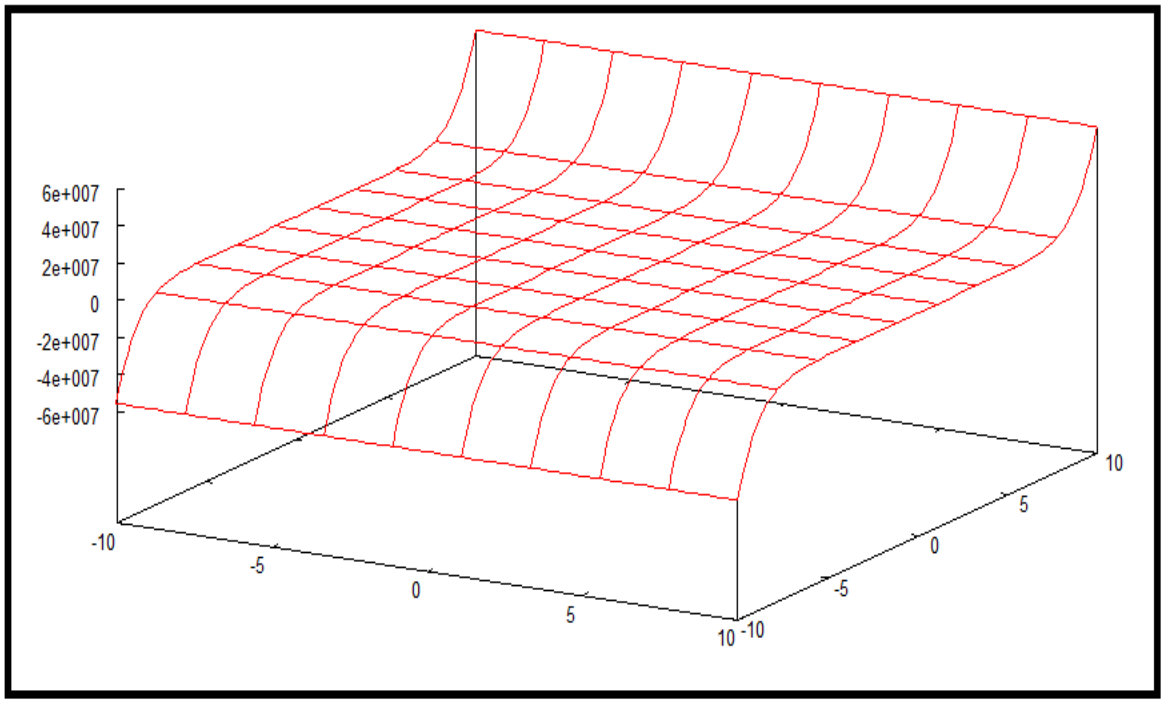

**Fig. 13:** Graph representing Kaniadakis entropy

### (vii) Geometry of Sharma- Mittal entropies

These are two-parameter families of entropic forms [26] known as Sharma- Mittal entropies. It can be written in the form:

$$S_{\kappa,r} = -\sum_i p_i^r \left( \frac{p_i^{\kappa} - p_i^{-\kappa}}{2\kappa} \right). \tag{30}$$

It is interesting to observe that they correspond to many other entropies as special cases. The Tsallis entropy is obtained for $r = \kappa$ and $q = 1 = -2\kappa$. The Kaniadakis entropy is obtained for $r = 0$. The Abe entropy is obtained for $\kappa = \frac{1}{2}(q - q^{-1})$ and $r = \frac{1}{2}(q + q^{-1}) - 1$. The Sharma-Mittal entropies are concave and Lesche stable. The geometrical plot of this entropic function appears as follows:

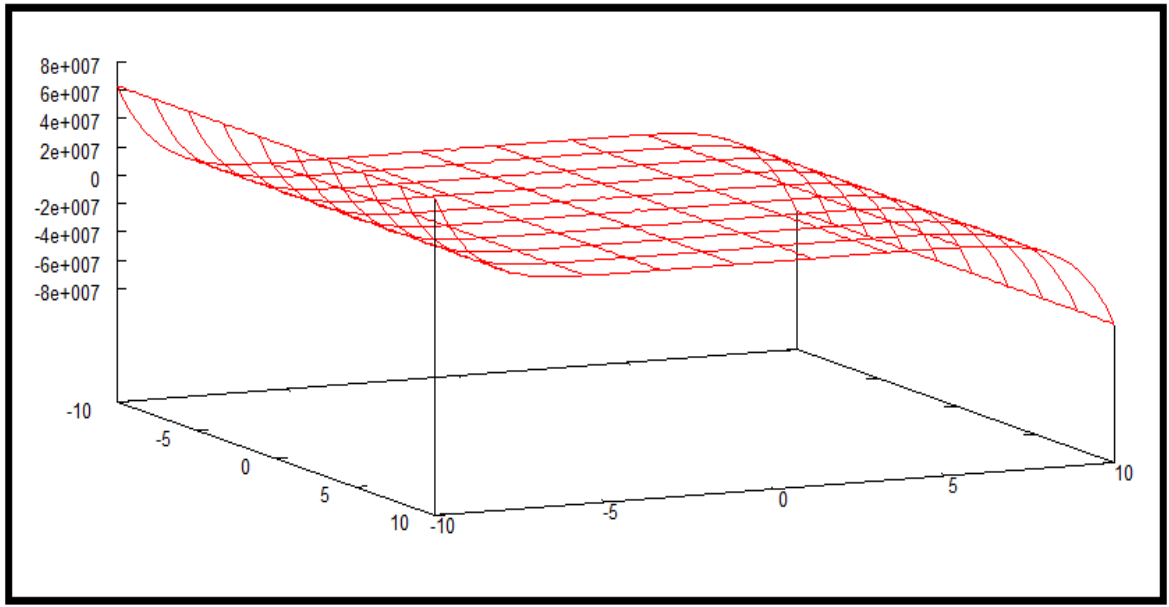

**Fig. 14:** Graph representing Sharma- Mittal entropy

## 4. Summary and Discussion

Having known about the loop character of the statistical geodesics in probability space, the geometry of the probability space in the form of *loop sequences* and *loop surfaces* is explored. The discussion in the present paper could prove to be of substantial importance in the ongoing research in Statistics and Mathematics, and also in Physics, and in particular in the studies of Gravity and Information theory. This also raises interesting philosophical questions such as: how two geodesics are manifested in two different geometrical forms, once when the probability of not happening of an event is included or else when this is not included. We arrive on remarkable conclusions that if we include the complementary probabilities $1 - p_1$ and $1 - p_2$ along with probabilities $p_1$ and $p_2$ in the information measure, we get more definitive picture in the form of a single loop only. And if we do not include the complementary probabilities, we can only have a loop sequence or a loop surface with vast extension of probabilities.

The geometry of Landsberg- Vedral entropy and Rényi entropy are pretty similar except that the two entropy functions hold at different scales. The geometry of Kaniadakis entropy and Sharma- Mittal entropy are pretty similar. The geometry of Abe entropy is similar to that of Tsallis entropy if it gets rid of the singularity at $q = 0$.

The stability conditions of different entropy functions could also be discussed in the light of this geometrical analysis. The geometries of Landsberg- Vedral entropy and Renyi entropy are pretty similar. Both of them have an unusual interface at $q = 0$ wherein two surfaces with opposite orientation meet. Probably this attribute could be seen as indicator that Landsberg- Vedral entropy and Renyi entropies are not Lesche stable.

We propose to study kinematical notions such as "mean free path" and "average scattering length" in the context of statistical distances and probabilities. And thus one could explore the concept of "mean statistical geodesics". Also, there is equally amusing question to be attended: Can the loops in probability space be multi-cornered as well?

† *The graphs in this paper have been plotted using Gnuplot.*

‡ *All graphs are plotted in 3-Dimenssional space. This goes well with the idea of the Holographic theory of entropy.*

# Acknowledgments

The author wishes to thank Prof. C. R. Rao, Prof K. R. Parthasarathi for their encouragement and suggestions.